# Effect of Nanoparticles on the Bulk Shear Viscosity of a Lung Surfactant Fluid

**L.P.A. Thai, F. Mousseau, E.K. Oikonomou, M. Radiom and J.-F. Berret***

*Matière et Systèmes Complexes, UMR 7057 CNRS Université Denis Diderot Paris-VII, Bâtiment Condorcet, 10 rue Alice Domon et Léonie Duquet, 75205 Paris, France.*

**Abstract** Inhaled nanoparticles (< 100 nm) reaching the deep lung region first interact with the pulmonary surfactant, a thin lipid film lining the alveolar epithelium. To date, most biophysical studies have focused on particle induced modifications of the film interfacial properties. In comparison, there is less work on the surfactant bulk properties, and on their changes upon particle exposure. Here we study the viscoelastic properties of a biomimetic pulmonary surfactant in the presence of various engineered nanoparticles. The microrheology technique used is based on the remote actuation of micron-sized wires *via* the application of a rotating magnetic field and on time-lapse optical microscopy. It is found that particles strongly interacting with lipid vesicles, such as cationic silica ($SiO_2$, 42 nm) and alumina ($Al_2O_3$, 40 nm) induce profound modifications of the surfactant flow properties, even at low concentrations. In particular, we find that silica causes fluidification, while alumina induces a liquid-to-soft solid transition. Both phenomena are described quantitatively and accounted for in the context of colloidal physics models. It is finally suggested that the structure and viscosity changes could impair the fluid reorganization and recirculation occurring during breathing.

# I - Introduction

Following their interaction with living organisms, nanoparticles (NPs) first interact with extracellular body fluids (ECF) before they come in contact with cellular structures. ECFs are composed of the blood plasma, of liquids from interstitial compartments, including the lymph, the mucus and the pulmonary surfactant and from small amounts of transcellular liquids in the ocular and cerebrospinal cavities. Combined, these fluids represent 1/3 of the total body fluids in humans, *i.e.* around 15 L for a person of normal weight. Upon contact with ECFs, NPs are dispersed in a waterborne phase rich in electrolytes, lipids, proteins and carbohydrates with which they interact. This is the case for example when NPs are introduced into the respiratory and digestive systems or the blood compartment. Over the last 20 years, studies have focused on NP interactions with serum proteins from the blood.[1,2] When mixed with serum, NP interfaces are spontaneously covered with proteins, leading to the protein corona formation.[3,4] For NPs, it is assumed that the protein corona determines their new biological identity and eventually regulate their interaction and biodistribution *in vivo*.

In the context of nanoparticles interacting with the ECFs, the respiratory zone in the lungs represents an interesting environment. In this region, the alveoli form a froth-like structure and their interface with the air is lined with a thin layer ($< 1 \mu m$) of pulmonary surfactant. Secreted by alveolar epithelial cells, pulmonary surfactant is an interstitial fluid that contains lipids and proteins in a 90:10 ratio, its overall weight concentration being approximately 40 g $L^{-1}$. Its role is to reduce the surface tension with the air, to prevent the alveoli collapse at the lowest lung volumes and facilitate their expansion during subsequent inspirations.[5,6] The proteins present in the surfactant film such as SP-A, SP-B, SP-C and SP-D play an important role in the structure and function of the alveoli, in particular with regard to pathogen elimination and to interfacial film stability. For particles not captured in the upper airways, surfactant represents the first physical barrier against small sized particles ($< 100$ nm). Pertaining to the interaction mechanisms with lipids, the picture emerging from current literature is based on the protein corona model described previously, leading to a description in terms of biomolecular or lipid coronas. For particles with hydrophilic surfaces, this corona is often depicted as a supported lipid bilayer (SLB) which consists in a single bilayer adsorbed at the particle surface *via* adhesive forces.[7-11] However, recent studies have shown that the SLB formation is not spontaneous (in contrast to the protein corona) and depends on various factors such as particle and membrane charges as well as particle size.[12-14] So far, the exact nature of this lipid corona is not well established.

In the alveolar spaces, inhaled NPs first cross the interfacial lipid monolayer and thereafter diffuse through a three-dimensional network of interconnected membranes towards the hypophase.[15] With a lipid concentration estimated at 40 g $L^{-1}$,[16-19] the hypophase is a dispersion where lipids are organized in multilamellar vesicles and tubular myelins.[6,19,20] To date, most biophysical studies have focused on the modification of the interfacial film properties. Using pulmonary surfactant mimetics, Schleh *et al.* have shown that nanosized titanium dioxide induces a surfactant dysfunction associated with a dose-dependent increase of adsorption and surface tension.[21] Similar results were found using gold, silica, polymeric NPs with either synthetic or exogenous formulations.[22-24] Surface pressure-area isotherms of synthetic lipid mixtures also depict strong modifications of the lipid interfacial organization caused by hydrophobic as well as hydrophilic NPs.[25-29] During compression and expansion, it is found that particles lead to a disruption of the lipid phase behavior and alter the surface tension hysteresis cycle. Transposed to the lung environment, repeated exposure is expected to cause perturbations of the lung physiology associated with particle retention, cellular oxidative stress generation and pro-inflammatory effects.[15,30,31]

In comparison, there is much less work done on the surfactant bulk properties and on the effects of nanoparticles. Bulk rheology measurements were reported by King *et al.*[32] and by Lu *et al.*[33] who found that the viscosity of native and exogenous surfactants determined at physiological concentrations were in the range $5 - 50$ mPa s, indicating a relatively low viscosity ECF. In a recent work, we investigated the viscosity of the biomimetic lung surfactant Curosurf® as a function of the volume fraction $\phi$ in vesicles and found that it obeys the well-known Krieger-Dougherty law for colloids,[34,35] $\eta(\phi) = \eta_S (1 - \phi/\phi_m)^{-2}$ where $\eta_S$ is the solvent viscosity and $\phi_m = 0.65$ the maximum-packing fraction.[36] The Krieger-Dougherty behavior is characterized by an exponential increase of the viscosity and a divergence at $\phi_m$. An important conclusion of this study was that in physiological conditions, the biomimetic lung fluid had a viscosity around 10 mPa s and a high-volume fraction ($\phi \sim 0.25$), indicating a crowded environment. With a physiological value close to the maximum-packing fraction, viscosity changes resulting from the addition of particles are expected. During breathing, pulmonary surfactant is subjected to local motion and any structure or viscosity changes could impair the fluid reorganization and recirculation.

In this work, we study the effects of silica and alumina nanoparticles on Curosurf® rheological properties at physiological concentrations. As NP models, silica and alumina were selected because they are manufactured in high volume by the chemical industry, which increases the risk of occupational and environmental exposure.[37] It was also found that silica and alumina NPs inhalation may lead to a series of disease including lung inflammation and cancer[38-40]. Recently, Puisney *et al.* have shown that silicon and aluminum based NPs, including oxides, are produced by brake wear and degradation processes of passenger vehicles and are expected to be present in urban air pollution in significant amounts.[41] Concerning the choices of nanoscale particles in our study, it has been shown that the particle deposition along the respiratory tract depends primarily on their aerodynamic diameter. It was found that for particles around 20 nm, more than 90% of the inhaled mass fraction deposit in the entire lungs and 50% in the alveolar region.[40,42] For the 40 nm particles put under scrutiny here, these percentages are slightly lower, at respectively 53% and 37%. For microrheology experiments, we exploit the technique of magnetic rotational spectroscopy (MRS)[43-46] in which micron-sized wires are submitted to a rotational magnetic field as a function of time and of the angular frequency ($\omega = 10^{-2} - 10$ rad s$^{-1}$), while their motion is monitored by time-lapse optical microscopy. The MRS technique enables to measure the shear viscosity and elastic modulus of complex fluids, it is suitable for fluids in minute amounts and with heterogeneities at the micron scale. It is found that particles interacting with Curosurf® vesicles induce profound modifications in the surfactant flow properties. At low concentrations, fluidification is observed with silica whilst with alumina, a transition towards an arrested vesicular state occurs at particle concentration above 0.1 g L$^{-1}$.

## II - Results and discussion

### II.1 - Nanoparticles, Curosurf® vesicles and interaction

Recently we studied the phase behavior of a series of nanoparticles and biomimetic surfactants under controlled physico-chemical conditions.[8,47,48] Interaction diagrams were determined from dilute solutions using various organic and inorganic NPs. From their aggregation behaviors with Curosurf®, the NPs were characterized with regard to their interaction strength parameter. Large strength parameters were obtained for highly charged cationic particles, whereas small or close to zero strength parameters were found for neutral and for anionic particles. In this section, we summarize the main results obtained on nanoparticle-vesicle interactions. Figs. 1a-b display transmission electron microscopy images of positively charged silica and alumina NPs, respectively. Silica are spherical and characterized by an average diameter of 42 nm, whereas alumina appear as irregular platelets of diameter 40 nm and thickness 10 nm. In solution, NP stability is ensured by electrostatic repulsions mediated by cationic surface charges. The NP charge density has been estimated at $+0.62e$ nm$^{-2}$ and $+7.3e$ nm$^{-2}$ respectively using a polymer-based titration method.[49] Combining different techniques such as the cryo-transmission electron microscopy (cryo-TEM), nanoparticle tracking analysis, dynamic light scattering and phase-contrast light microscopy, we have shown that Curosurf® lipids self-assemble into vesicles and that the vesicle sizes are broadly distributed, typically between 100 nm and 10 μm.[36] Fig. 1c illustrates a representative cryo-TEM image of Curosurf® at 5 g L$^{-1}$ showing uni-, multilamellar and multivesicular vesicles.[8,21,50-53] From the lipids in Curosurf® (Supplementary Information S1), we determined the surface charge density on the vesicles at -0.67 $e$ nm$^{-2}$ and anticipate strong (resp. weak) interactions with cationic (resp. anionic) NPs.[49,54]

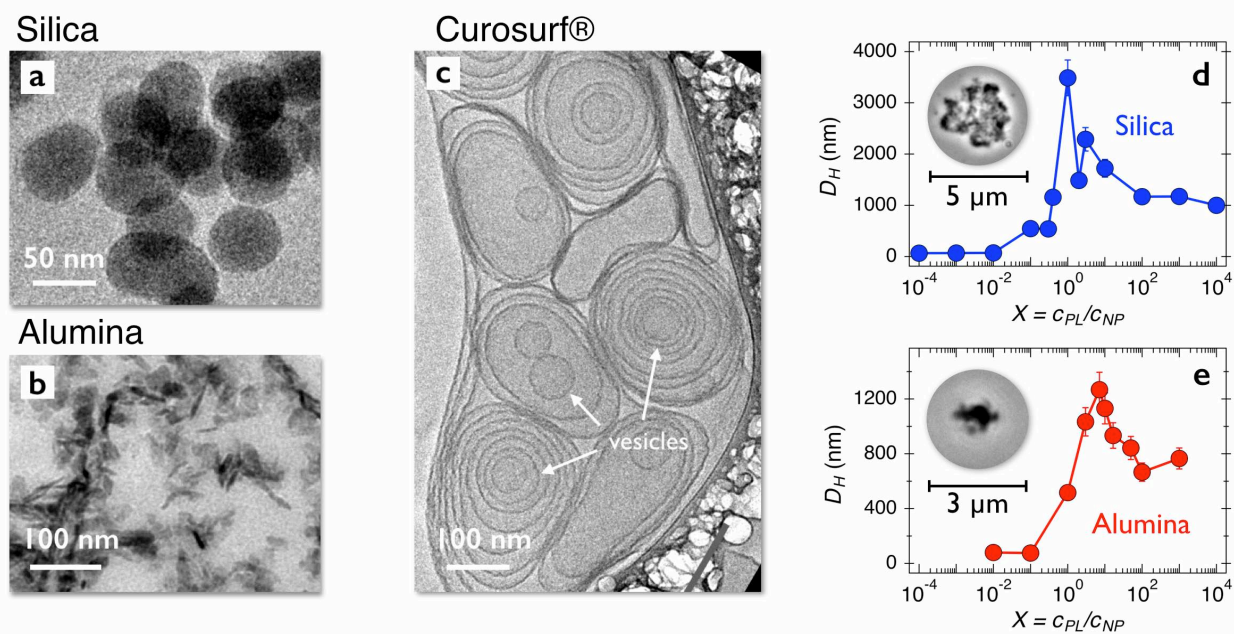

**Figure 1: a, b)** *Transmission electron microscopy (TEM) images of positively charged silica and alumina nanoparticles. Their sizes are respectively 42 nm and 40 nm (platelet diameter) and their size dispersities (ratio between the standard deviation and the mean) are s = 0.11 and 0.30.* **c)** *Cryogenic TEM images of native Curosurf® obtained from a 5 g L$^{-1}$ dispersion at 25 °C. The arrows are pointing to multi-lamellar vesicles and multi-vesicular vesicles. Additional images are available in Supplementary Information S4.* **d, e)** *Hydrodynamic diameter of nanoparticle-vesicle dispersions as a function of the concentration ratio X = $c_{PL}/c_{NP}$ obtained from dynamic light scattering at T = 25 °C for silica (d) and for alumina (e). In the figures the lipid and nanoparticle weight concentrations as a function of X read: $c_{PL}(X) = Xc/(1 + X)$ and $c_{NP}(X) = c/(1 + X)$, with c being the total concentration. To avoid multiple scattering, the total active concentration was adjusted at c = 1 g L$^{-1}$ for silica and c = 0.1 g L$^{-1}$ for alumina.* **Inset**: *Optical microscopy images of mixed nanoparticle-vesicle aggregates. The images were obtained from dispersions at X's corresponding to the maximum seen by light scattering.*

To evaluate these interactions, we use the method of continuous variation developed by P. Job that we modify for light scattering.[47,55,56] Figs. 1d and 1e display the hydrodynamic diameter obtained from mixed dispersions as a function of the concentration ratio $X = c_{PL}/c_{NP}$, where $c_{PL}$ and $c_{NP}$ denote the lipid and nanoparticle weight concentrations at a given $X$. Note that in the Job plots the total concentration $c = c_{PL} + c_{NP}$ is held constant and maintained in the dilute regime. In Fig. 1d and 1e, the hydrodynamic diameters exhibit a maximum in the intermediate range $X = 1 - 10$, which we ascribe to the formation of aggregates. Examples of aggregates seen in optical microscopy are shown in the insets. Results on aggregates were obtained at room and body temperatures *i.e.* below and above the gel-to-fluid transition of Curosurf® membrane at 29.5 °C (Supplementary Information S2).[8,48] The outcomes of Fig. 1 suggest a strong electrostatic attraction between the oppositely charged NPs and vesicles. According to this scenario, the NPs adsorb at the lipid membranes and play the role of stickers for vesicles. Instances of particle sticking at the vesicular membrane were found with silica NPs using cryo-TEM experiments.[47] As a negative control, we also investigated the effects of an anionic silica NPs (20 nm) with Curosurf®. The Job scattering plot did not show evidence of aggregation and the $D_H$'s varied continuously from that of the NPs to that of the vesicles. Results on anionic silica NPs and interaction diagrams are provided in Supplementary Information S3.

## II.2 - Nanoparticle loaded surfactant showing a viscous fluid behavior

We now turn to the study of Curosurf® and to the effects of NPs on the rheology. The Curosurf® concentration is fixed at $c_{PL} = 44$ g L$^{-1}$, while the NP concentrations are varied from $c_{NP} = 10^{-3}$ to 0.5 g L$^{-1}$. It is shown in Materials and Methods that the lowest range of the above $c_{NP}$-interval is associated with an amount of nanoparticles in the alveolar region (extrapolated to human lungs) of 20 to 200 μg.[21,42] Depending on the NP type and concentration, two generic behaviors were observed with the wire-based microrheology. In this section, we describe the features corresponding to the case where the wires rotate steadily in the fluid with a non-zero average velocity, defined as $\Omega(\omega) = < d\theta(\omega,t)/dt >_t$, $\theta(\omega,t)$ being the wire orientation angle at actuating frequency $\omega$. The case where $\Omega(\omega) \simeq 0$ is discussed in the next section. Fig. 2a illustrates the rotation of a 43 μm wire in a Curosurf® dispersion mixed with alumina NPs. The different images of the chrono-photograph are taken at fixed time intervals (3.5 s) during a $\pi$-rotation of the object, showing that the wire rotates synchronously with the field. The applied frequency is 0.06 rad s$^{-1}$ and the magnetic field $\mu_0 H = 10.3$ mT. In the image background, micron-sized vesicles may be seen and in Movie#1&2, it is shown that they are being sheared by the wire rotation (Supplementary Information, Movies#1&2). With increasing frequencies, the wire exhibits a transition between a synchronous and an asynchronous (*i.e.* not synced with the magnetic excitation) regime. In Figs. 2b, 2c and 2d, transient $\theta(t)$-traces are represented at angular frequencies 0.06, 0.3 and 3 rad s$^{-1}$. The straight lines in red are least squared fitting to the $\theta(t)$ in both regimes, and their slopes provide the average rotation velocity $\Omega$, here equal to 0.06, 0.12 and 0.03 rad s$^{-1}$. In these studies, non-zero $\Omega(\omega)$'s were found for pristine Curosurf® at physiological concentrations[36] and for Curosurf® mixed with silica at all concentrations tested, and with alumina at concentrations below $c_{NP} = 0.1$ g L$^{-1}$. Such behaviors are characteristic of viscous or viscoelastic liquids.[57]

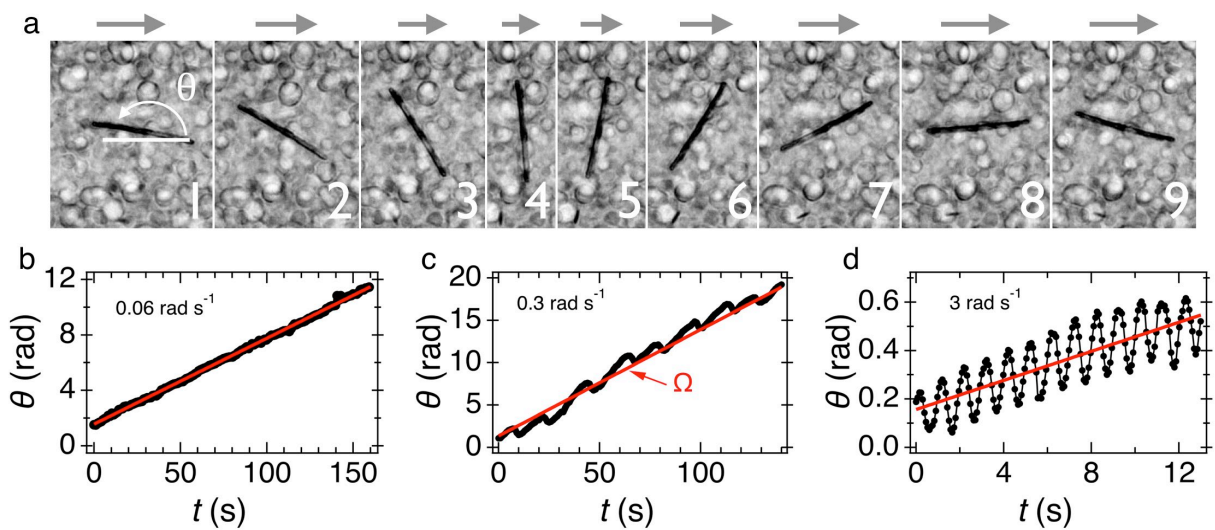

***Figure 2: a)*** *Chronophotograph of a 43 μm wire undergoing a π-rotation (angular frequency 0.06 rad s⁻¹, magnetic field μ₀H = 10.3 mT) in a 44 g L⁻¹ pulmonary surfactant dispersion loaded with alumina NPs ($c_{NP}$ = 0.004 g L⁻¹). The time interval between two images is 3.5 s. The upper arrows indicate the direction of rotation (here clockwise).* ***b)*** *Time dependence of the rotation angle $\theta(t)$ in the synchronous regime ($\omega$ = 0.06 rad s⁻¹) observed in the sample of Fig. 2a.* ***c and d)*** *Same as in Fig. 2b for the asynchronous regime at $\omega$ = 0.3 and 3 rad s⁻¹. The average rotation frequency $\Omega(\omega)$ is defined from the straight lines indicated in red.*

The rheological nature of the vesicular fluids put under scrutiny is here identified by monitoring the asymptotic behaviors of two measurable quantities: the angular frequency $\omega_C$ at which the synchronous-asynchronous transition takes place, and the angle $\theta_B(\omega)$ which is the amplitude of the oscillations in the asynchronous regime. As shown previously,[46,57,58] $\omega_C$ is used to determine the fluid viscosity $\eta$ through the expression[57] $\omega_C = 3\mu_0\Delta\chi H^2/8\eta L^{*2}$ where $L^* = L/D\sqrt{g(L/D)}$ and $g(L/D) = ln(L/D) - 0.662 + 0.917D/L - 0.050(D/L)^2$. In the previous equation, $\mu_0$ is the permeability in vacuum, $L$ and $D$ the length and diameter of the wire, $H$ the magnetic excitation amplitude, $\Delta\chi = \chi^2/(2+\chi)$ denotes the anisotropy of susceptibility between parallel and perpendicular directions and $\chi$ represents the material magnetic susceptibility. Fig. 3a displays the ratio $8\eta\omega_C/3\mu_0\Delta\chi H^2$ as a function of $L^*$ for different Curosurf® samples with and without NPs. The definition for $\omega_C$ is illustrated in the inset. The concentrations shown in the figure are $c_{NP}$ = 0.037 and 0.086 g L⁻¹ for alumina and $c_{NP}$ = 0.088, 0.21 and 0.50 g L⁻¹ for silica. The data points are found to collapse on a single master curve displaying the $1/L^{*2}$-dependence (continuous line). The agreement between data and theory is excellent. The measured viscosity is between 3 and 12 mPa s and its concentration dependence will be discussed in Section. III.4. Fig. 3b shows the oscillation amplitude angle $\theta_B$ as a function of the reduced frequency $\omega/\omega_C$ for the same 6 samples (see inset for the $\theta_B$-definition). A good superimposition of the data points is observed and is in agreement with the Newtonian constitutive equation prediction which goes as $\theta_B(\omega/\omega_C) \sim (\omega/\omega_C)^{-1}$ for $\omega \gg \omega_C$ (continuous line).[57,58] Note that in the $\theta_B(\omega/\omega_C)$-representation, the continuous line in the figure is obtained with no adjustable parameter.[45,57] In conclusion to this part, it is found that Curosurf® loaded with silica behaves as a Newtonian fluid at all concentrations studied and that the dispersions are characterized by a single rheological parameter, the static shear viscosity $\eta$. A similar type of behavior can be observed for Curosurf® mixed with alumina at concentrations lower than $c_{NP} = 0.1$ g L$^{-1}$. In the previous samples, evidence of viscoelasticity was not found.[36,57]

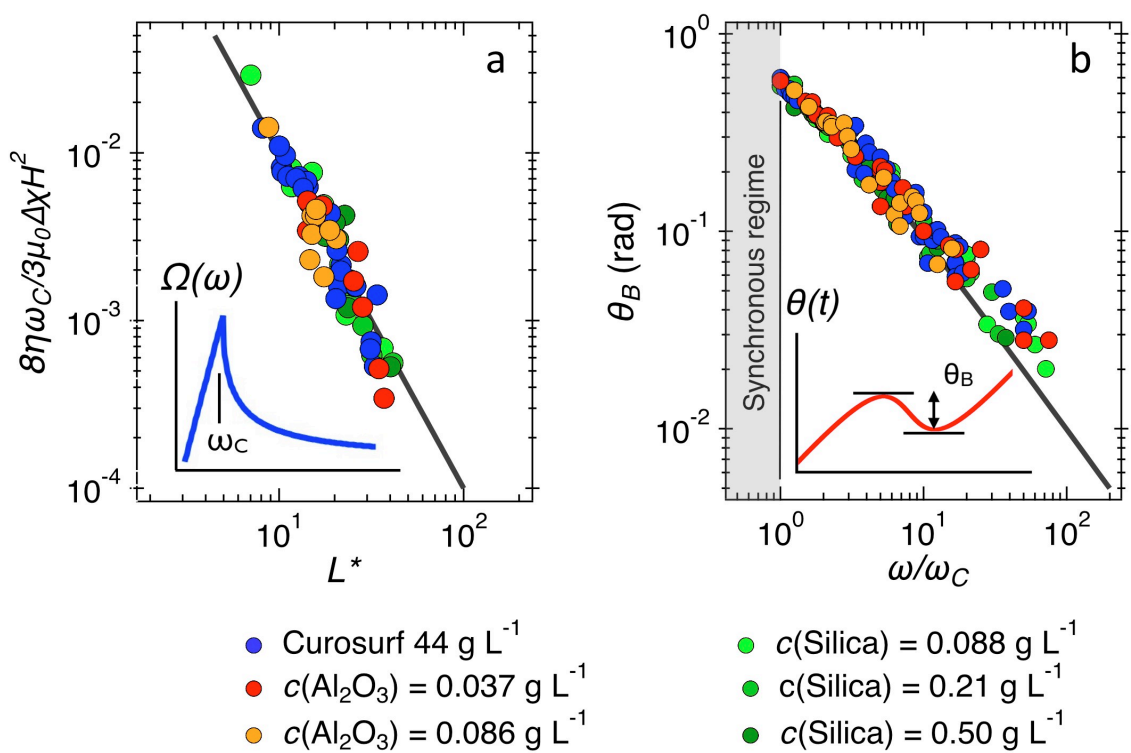

**Figure 3: a)** Normalized critical frequency $8\eta\omega_C/3\mu_0\Delta\chi H^2$ as a function of the reduced wire length $L^*$ obtained for Curosurf® dispersions with and without particles (T = 25 °C). The Curosurf® concentration is set at the physiological value, 44 g L$^{-1}$ and the nanoparticle concentrations are $c_{NP} = 0.037$ to 0.086 g L$^{-1}$ for alumina and 0.088 to 0.50 g L$^{-1}$ for silica. The parameters featuring in the normalized critical frequency are given in the text. The straight line displays the $1/L^{*2}$-dependence predicted from the viscous fluid constitutive equation. **Inset**: Angular frequency dependence of the average rotation velocity $\Omega(\omega)$ illustrating the definition of $\omega_C$. **b)** Oscillation amplitude $\theta_B(\omega/\omega_C)$ observed in the asynchronous regime for the same samples as in **a)** The straight line is for Newtonian fluids. **Inset**: Time-dependent rotation angle of a wire in the asynchronous regime illustrating the definition of $\theta_B$.

## II.3 - Alumina loaded surfactant showing a soft solid behavior

Here we describe experiments corresponding to the second generic behavior in which wires oscillate apart from a fixed orientation and have a zero-average velocity. Fig. 4a displays a chronophotograph of a 64 μm wire incorporated to a 44 g L$^{-1}$ Curosurf® dispersion mixed with alumina particles at $c_{NP} = 0.40$ g L$^{-1}$ ($\omega = 0.3$ rad s$^{-1}$, $\mu_0H = 10.3$ mT). The corresponding movie can be watched in Supplementary Information, Movies#3&4. The images in Fig. 4a are recorded every 3.5 s and show that the wire now oscillates between two orientations, here $\theta = 85°$ (as shown in images 1, 4 and 7) and $\theta = 55°$ (images 2, 5 and 8). These back-and-forth oscillations remain steady over long period of time (> 1 h). The top red arrows indicate the counterclockwise rotation occurring at each oscillation. Fig. 4b, 4c and 4d display the orientation angle $\theta(t)$ at angular frequencies 0.06, 0.3 and 3 rad s⁻¹ respectively. As shown by the straight lines in red, the average rotation frequency $\Omega(\omega)$ is small (of the order of $\pm 10^{-4}$ rad s⁻¹) and frequency independent. This inequality $\Omega(\omega)/\omega \ll 1$ observed in these experiments is an indication that the material around the wires is a soft solid and characterized by a yield stress behavior. The results of Fig. 4 are similar to those found in calcium ion crosslinked polysaccharide gels and for which cone-and-plate rheometry evidenced a marked yield stress signature.[45]

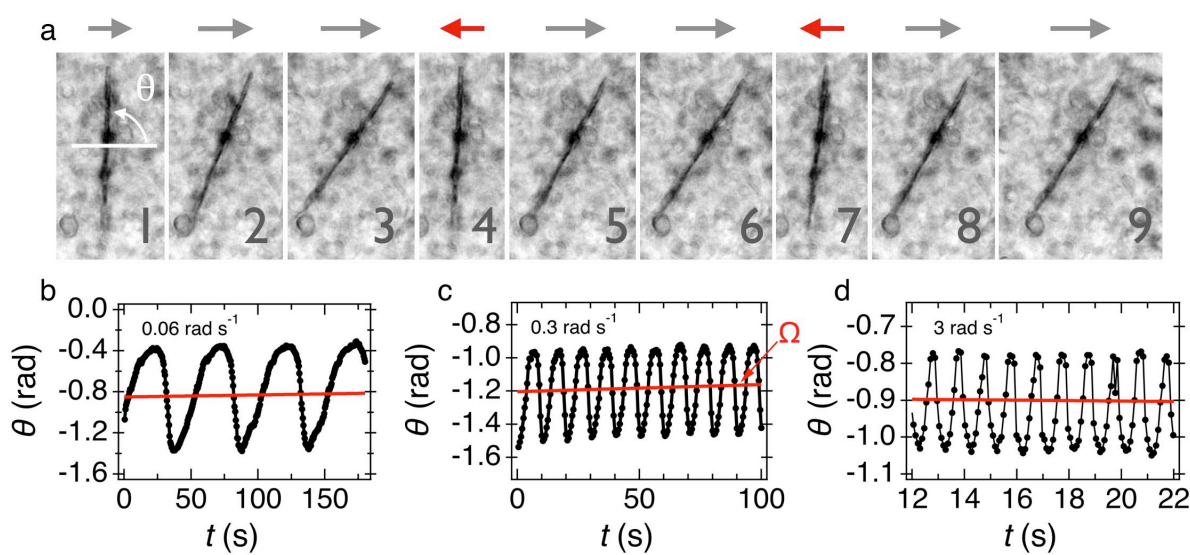

***Figure 4: a)** Chronophotograph of a 64 µm wire undergoing back-and-forth oscillations in the mimetic pulmonary surfactant Curosurf® loaded with alumina NPs ($c_{NP} = 0.40$ g L⁻¹). The angular frequency is set at 0.3 rad s⁻¹ and the magnetic field at 10.3 mT. Time interval between two images is 3.5 s. The upper grey (resp. red) arrows indicate a clockwise (resp. anticlockwise) movement of the wire. **b, c and d)** Time dependences of the rotation angle $\theta(t)$ in the sample shown in **a)** at $\omega = 0.06$, 0.3 and 3 rad s⁻¹, respectively. The average rotation frequency $\Omega(\omega)$ is defined from the straight lines indicated in red. Movies corresponding to Figs. 4b and 4d are in Supplementary Information, Movies#3&4).*

We now proceed to a quantitative analysis of the data obtained on the soft solid dispersions. Fig. 5a displays the oscillation amplitude $\theta_B(\omega)$ as a function of the frequency at $c_{NP} = 0.19$ g L⁻¹ and for wires comprised between 28 and 92 µm. It is found that $\theta_B(\omega)$ exhibits a plateau at low frequency, noted $\theta_{B,Max}$ and decreases above 1 rad s⁻¹. At $c_{NP} = 0.49$ g L⁻¹, the $\theta_B(\omega)$-response consists essentially in a plateau and no high frequency decrease is observed (Supplementary Information S5). According to the constitutive equations developed for soft solids,[45] the low frequency limit, here approximated to $\theta_{B,Max}$ is linked to the equilibrium storage modulus $G_{eq}$ through the relation $\lim_{\omega \to 0} \theta_B(\omega) = 3\mu_0 \Delta \chi H^2 / 4L^{*2} G_{eq}$. For a soft solid, $G_{eq}$ characterizes the quasi-static elastic response of the material, i.e. as $\omega$ goes to zero. Figs. 5b and 5c show the $L^*$-dependence of the maximum angle for two alumina concentrations together with the predicted $\theta_{B,Max} \sim 1/L^{*2}$ scaling.[45] The shaded area in the figures highlights the range over which the $1/L^{*2}$-prefactor varies in these experiments. These results suggest that the variability in the modulus could come from elasticity fluctuations at the scale of the wires. The value and physical interpretation of the elastic soft solid properties are discussed in the next section.

To further illustrate the soft solid behavior found for $Al_2O_3$ loaded dispersions, we investigated the transmitted light intensity with time-lapse microscopy in pristine and in processed samples. From their structures, the samples with and without NPs appear identical under the phase-contrast microscope (Fig. 5d). Strong differences are observed however with regards to the transmitted intensity light fluctuations (Fig. 5e). For Curosurf®, the vesicles are agitated with thermal motions, leading to noticeable fluctuations of the transmitted light. For $Al_2O_3$-loaded samples by contrast, the vesicles appear as stuck to each other and form large immobile clusters (Supplementary Information S6 and Movie#5). The observed vesicular frozen state is illustrated in Fig. 5e, which compares the fluctuation distributions for Curosurf® dispersions with and without alumina particles ($c_{NP} = 0.50$ g L$^{-1}$). The distributions were collected at five different locations on each sample by recording the transmitted light intensity as a function of the time and were later adjusted with a Gaussian function. Best fit calculations reveal that the fluctuation amplitudes are decreased by a factor of 10 for alumina compared to those of the initial dispersion and that the data for silica are close to those of pristine dispersion. These results illustrate quantitatively the existence of an arrested state induced by alumina particles, in agreement with the microrheology findings. In conclusion, we found that concentrated alumina loaded Curosurf® behaves as a soft solid material characterized by a yield stress and by an equilibrium storage modulus $G_{eq}$, the latter being measured. For these samples, the static viscosity $\eta$ is infinite.[35]

## II.4 - Concentration dependences

Fig. 6 shows the concentration dependences of the rheological properties of Curosurf® loaded with silica and alumina NPs. The main frame displays the static viscosity $\eta$ as a function of the nanoparticle concentration $c_{NP}$ (data from Section III.2). Below 10$^{-3}$ g L$^{-1}$, the viscosity remains at the level of the pristine fluid around $\eta = 9.0 \pm 2.0$ mPa s. This value agrees well with the recent measurements obtained on Curosurf® as a function of the lipid content.[36] With increasing $c_{NP}$, a similar behavior is observed with the two particles studied. The static viscosity decreases down to $5 \pm 1$ mPa s and for silica the tendency persists towards the highest concentrations. At 0.5 g L$^{-1}$, $\eta(c_{NP})$ has been reduced by a third, eventually reaching $3.5 \pm 0.7$ mPa s. The straight lines in the figure are power laws of the form $\eta(c_{NP}) \sim c_{NP}^{-0.15}$, indicating that the viscosity fall is slow and progressive. Note that for the non-interacting anionic silica, the viscosity remains concentration independent (Supplementary Information S7).

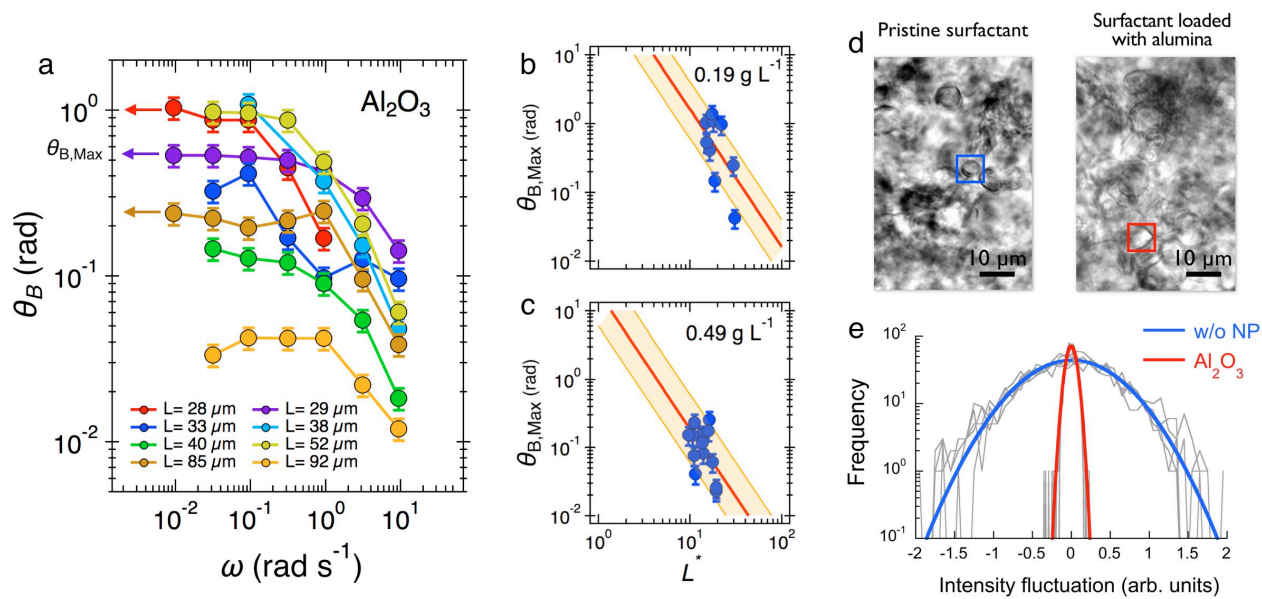

**Figure 5: a)** *Oscillation amplitude $\theta_B(\omega)$ measured in Curosurf® loaded with alumina particles ($c_{NP} = 0.19$ g $L^{-1}$) using wires of different lengths. The arrows at low frequencies indicate the maximum values $\theta_{B,Max}$ detected in frequency sweeps.* **b, c)** *$\theta_{B,Max}(L^*)$ for Curosurf® dispersions loaded with alumina particles at $c_{NP} = 0.19$ and $0.49$ g $L^{-1}$ respectively, $L^*$ being the reduced wire length defined in the text. The thick red line indicates the $1/L^{*2}$-behavior expected from the soft solid model.[45] The orange areas denote the range over which the $1/L^{*2}$-prefactor varies in these experiments. These ranges are associated with a standard error of 25%.* **d)** *Phase contrast microscopy images of Curosurf® samples without and with alumina particles. The squares indicate spatial domains where the transmitted light intensity was collected as a function of the time.* **e)** *Distribution of the transmitted light fluctuation amplitudes for Curosurf® and Curosurf® treated with alumina NPs ($c_{NP} = 0.50$ g $L^{-1}$). The data in grey are distributions taken in 5 different locations, and the continuous curves are best fit calculations using a Gaussian function. The standard deviations for the continuous lines in blue and red are 0.38 and 0.05, respectively.*

The reduction of viscosity noted in Fig. 6 is a phenomenon that can be understood in the context of colloidal physics. Recent studies have found that for dense suspensions stabilized by short range repulsion (as it is the case for Curosurf® vesicles), the addition of weak attractive interaction leads to a decrease in viscosity.[59-62] In these reports, the colloids are beads and microgels and the attraction is due to depleting non-adsorbing polymers.[59,61,62] It was found that a minute amount of depleting agents leads to the formation of large cohesive aggregates and thereby to a decrease of the effective volume fraction. As a result, the suspension bulk viscosity decreases, in proportions similar to those of Fig. 6.[60,61] With NPs strongly interacting with vesicles (as shown in Fig. 1d and 1e), results suggest that the attractive potential is not due here to depletion forces, but mediated by particles acting as physical links between vesicles.[8,48] However, in the low concentration regime ($c_{NP} < 0.1$ g $L^{-1}$), the particle shape and surface charge density do not play a major role, as the viscosities of silica and alumina loaded dispersion behave similarly.

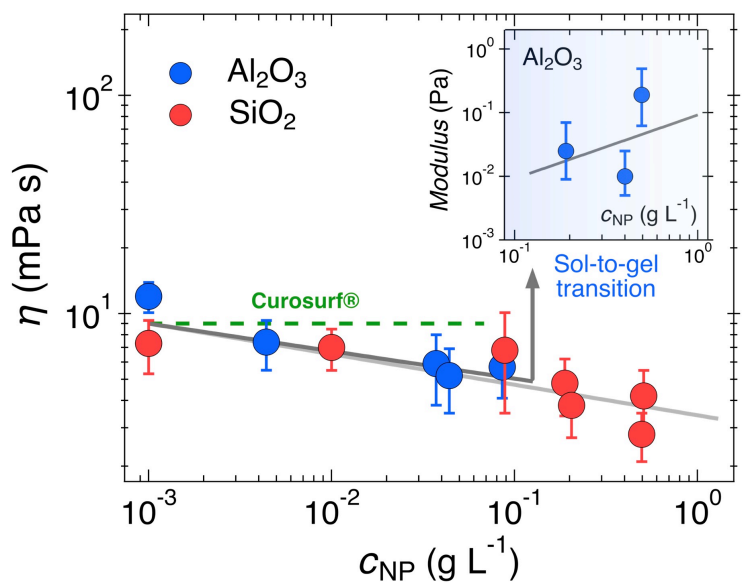

**Figure 6:** *Static shear viscosity of Curosurf® loaded with silica (red symbols) and alumina (blue symbols) as a function of the nanoparticle concentration $c_{NP}$. The straight lines in grey are least squared calculations using power laws of the form $\eta(c_{NP}) \sim c_{NP}^{-\alpha}$ with $\alpha \sim 0.15$. The vertical arrow shows the transition between a liquid and a soft solid behavior. **Inset**: Equilibrium elastic modulus $G_{eq}$ for alumina loaded surfactant above $c_{NP} = 0.1$ g $L^{-1}$. The straight line is calculated from the concentration dependence of the instantaneous elastic modulus assuming that all alumina particles are active bonds in the percolated vesicular network.*

The second important result is the observation of a transition from a viscous liquid to a soft solid for alumina dispersions (inset of Fig. 6). The measured equilibrium modulus $G_{eq}(c_{NP})$ is found in the range $10^{-2} - 10^{-1}$ Pa and increases with the concentration. At such values, the vesicle network is fragile and can be easily broken *e.g.* by the application of mild flow conditions. To explain the soft solid formation, we again exploit the colloid analogy and assume that with strongly interacting NPs, the vesicle dispersion undergoes a bond percolation transition. Such transitions are common in liquid-to-soft solid phenomena involving polymers or colloids.[35] To substantiate this assertion, we estimate the instantaneous elastic modulus $G_0(c_{NP})$ by assuming that all alumina particles added in the dispersion are active bonds and participate to the cross-linked network. In this case, $G_0(c_{NP})$ scales with the product $n_{NP}k_BT$, where $n_{NP}$ is the bond density, $k_B$ the Boltzmann constant and $T$ the absolute temperature.[35] In regular soft solid, the instantaneous and equilibrium moduli are in principle different, but an estimation for $G_0(c_{NP})$ is sometimes useful to impart orders of magnitude for the elastic response. The straight line in the inset displays the linear behavior predicted for $G_0(c_{NP})$ and provides moduli in the range $10^{-2} - 10^{-1}$ Pa, as for the experiments. Concerning the soft solid properties, it is possible that other mechanisms such as crowding or modification of the membrane elasticity contribute to the overall dispersion elasticity. The differences between silica and alumina particles are explained by the charge and shape of the particles. The silicas are spherical and weakly charged, whereas alumina NPs are in the form of platelets and strongly charged. Cryo-TEM images have also shown that the silica particles can be internalized in the vesicles, potentially reducing the number of effective bonds in the network formation.[47] In conclusion, it is found that above a threshold concentration, the pulmonary surfactant Curosurf® loaded with alumina particles undergoes a rheological transition from a liquid to a soft solid state. These variations in viscosity are important for alveoli whose function is based on the flow properties of the surfactant layer, as they could cause pulmonary diseases.

## III - Conclusion

In this work we study the rheological properties of the biomimetic pulmonary surfactant Curosurf® in the presence of engineered nanoparticles. We use a recent microrheology technique developed in-house and based on the remote actuation of micron-sized wires *via* the application of a rotating magnetic field. The wire microrheology is quantitative as it can measure the static shear viscosity and storage modulus of complex fluids. It also enables to differentiate between fluid-type and soft solid-type behaviors,[57] an outcome that is often hard to achieve with classical cone-and-plate rheometers. In a recent work we investigated the viscosity of Curosurf® dispersions as a function of the lipid concentration, and we found that in physiological conditions, the vesicle volume fraction is high and close to the liquid-to-soft solid transition. This finding prompted us to examine the effect of NPs on the surfactant rheology. Here, we extend the work on this biomimetic lung fluid and study its rheological behavior in presence of sub-100 nm NPs. Three types of particles are put under scrutiny, silica particles bearing either positive or negative charges at their surface and positively charged alumina. For particles strongly interacting with lipid membranes *via* electrostatic attraction, such as positive silica and alumina, significant changes to the dispersion viscosity are observed. For the positive silica, a reduction of the viscosity by a factor of 3 is found from $\eta = 9.0 \pm 2.0$ mPa s for pristine Curosurf® to $3.5 \pm 0.7$ mPa s at $c_{NP}$ = 0.5 g L$^{-1}$. For alumina, we observe a liquid-to-soft solid transition at $c_{NP}$ = 0.1 g L$^{-1}$ that is interpreted in terms of a bond percolation transition, the particles acting as cross-links in regards to the vesicular network. These results show that inhaled particles interacting with lipid membrane can significantly alter the surfactant flow properties and vesicle reorganization taking place during breathing, a phenomenon that has not been evaluated yet.

## IV - Materials and Methods

### IV.1 - Materials

*Nanoparticles*

Aluminum ($Al_2O_3$) and silicon oxide ($SiO_2$) nanoparticle synthesis and characterization have been described recently.[8,47-49] The NPs particles were chosen based on their importance in the nano-material production worldwide and on the inherent risks of inhalation exposure. A list of the production volumes and industrial applications for these NPs can be found in Supplementary Information S8. Positive silica were synthetized *via* the Stöber synthesis using tetraethyl orthosilicate silica precursor (TEOS, Aldrich).[8] Functionalization by amine groups was carried out, resulting in a positively charged coating. Dispersions were prepared at 40 g L$^{-1}$ and diluted with DI-water at pH 5 for further use. The geometric diameter was determined from TEM at $D_0 = 41.2$ nm (dispersity 0.11, Fig. 1a). Negative silica (trade name CLX®, diameter 20 nm, dispersity 0.20) were purchased from Sigma Aldrich at the concentration of 450 g L$^{-1}$. The batch was diluted down to 50 g L$^{-1}$ and dialyzed against DI-water at pH 6.4 for 48 h. Alumina dispersions were prepared from Disperal® (SASOL, Germany) powder at the concentration of 10 g L$^{-1}$ (pH 5) and thoroughly sonicated. TEM images show irregular platelets of length 40 nm and thickness 10 nm, the dispersity being $s = 0.30$ (Fig. 1b). The dispersion stability was assessed using dynamic light scattering at pH 5 and 6.4 over extended periods of time (> months)[47] and the hydrodynamic diameters were found to be time independent at $D_H = 64$ nm, 60 nm and 34 nm, respectively. The surface charge density was determined using the polyelectrolyte assisted charge titration spectrometry at $\sigma = +7.3e$, $+0.62e$, and $-0.31e$ nm$^{-2}$ (Table I).[49]

| Nanoparticle | $D_0$ (nm) | $s$ | $D_H$ (nm) | $\sigma$ (nm$^{-2}$) |
|---|---|---|---|---|
| Alumina | 40 | 0.30 | 64 | $+7.3e$ |
| Silica (positive) | 42 | 0.11 | 60 | $+0.62e$ |
| Silica (negative) | 20 | 0.20 | 34 | $-0.31e$ |

*Table I*: List of nanoparticles and their characteristics. $D_0$ and $D_H$ stand for the geometric and hydrodynamic diameters determined by transmission electron microscopy and dynamic light scattering. $s$ denotes the size dispersity (ratio between the standard deviation and average size of the distribution). The electrostatic charge densities $\sigma$ was obtained using polyelectrolyte assisted charge titration spectrometry.[49]

*Pulmonary surfactant*

Curosurf® (*Chiesi Pharmaceuticals*, Parma, Italy) is a porcine minced pulmonary surfactant extract used in maternity hospitals for the treatment of premature newborns with respiratory distress syndrome.[63] It is produced as an 80 g L$^{-1}$ whitish dispersion at pH 6.4 and contains phosphatidylcholine (PC) lipids, sphingomyelin (SM), phosphatidylethanolamine (PE), phosphatidylinositol (PI), phosphatidylglycerol (PG) and the hydrophobic proteins SP-B and SP-C.[50,64,65] Curosurf® is considered as a model biomimetic surfactant because its composition and bulk properties are constant from one batch to another and because of its long-term shelf stability. The Curosurf® formulation is compared to pulmonary surfactant obtained from lung lavage in Supplementary Information S1. The main differences between the two formulations are the percentages in sphingomyelin, in cholesterol and in phosphatidylglycerol, as well as the absence of the hydrophilic proteins, SP-A and SP-D. 75 μl aliquots were frozen, stored at -20 °C and thawed before use. Curosurf® was kindly provided by Dr. Mostafa Mokhtari and his team from the neonatal service at Hospital Kremlin-Bicêtre, Val-de-Marne, France and used as received or diluted in phosphate buffer saline (PBS, Aldrich). Recently, we used the microrheology MRS technique described below to measure Curosurf® viscosity as a function of the volume fraction[36] and observed an exponential increase in the range $0 - 80$ g L$^{-1}$. The $\eta(\phi)$-data are provided in Supplementary Information S9.

*Sample preparation*

For the evaluation of the interaction strength parameters (data from Fig. 1), dilute stock solutions of surfactant and NPs were prepared in the same conditions of pH and concentration. The dispersions were then mixed at different ratios $X = c_{PL}/c_{NP}$ following the continuous variation protocol.[47,55,56] The total concentration was kept constant at $c = c_{PL} + c_{NP} = 0.1$ or $1$ g L$^{-1}$, *i.e.* below the lipid physiological concentration. For alumina, the pH of the stock solution was adjusted at pH 5 to ensure that particles do not aggregate as a result of pH changes. Positive and negative silica were studied at the surfactant physiological pH. For microrheology experiments, the *Chiesi* formulation at $c_{PL} = 80$ g L$^{-1}$ was used as received and diluted by addition of a NP dispersion in PBS, ensuring physiological conditions of pH and ionic strength. In the end, the lipid concentration was at $c_{PL} = 44$ g L$^{-1}$ and that of the nanoparticles was varied from $c_{NP} = 10^{-3}$ to $0.50$ g L$^{-1}$. From these values, it is possible to derive the total mass present in the alveolar region, a quantity that is of interest for the comparison with actual exposure data. Assuming for the human alveolar region a total volume of pulmonary surfactant of 25 mL,[19,21] we found that the lowest concentration used, $c_{NP} = 10^{-3}$ g L$^{-1}$ corresponds to 24 µg of NPs in the entire alveolar region, and that the first effect on the viscosity, observed around $10^{-2}$ g L$^{-1}$ corresponds to 240 µg. These two values are associated with a total number of particles per alveolus of 1000 and 10000 respectively. Detailed calculations, including those for alumina and negative silica are given in Supplementary Information S10.

## IV.2 - Methods

*Optical and cryo-transmission electron microscopy*

Phase-contrast and bright field images were acquired on an IX73 inverted microscope (Olympus) equipped with 20× and 100× objectives. An EXi Blue camera (QImaging) and Metaview software (Universal Imaging Inc.) were used as acquisition system. For cryo-TEM, few microliters of a $c_{PL} = 5$ g L$^{-1}$ Curosurf® dispersion were deposited on a lacey carbon coated 200 mesh (Ted Pella Inc.). The drop was blotted with a filter paper using a FEI Vitrobot™ freeze plunger. The grid was then quenched rapidly in liquid ethane to avoid crystallization and later cooled with liquid nitrogen.[66] The membrane was transferred into the vacuum column of a JEOL 1400 TEM microscope (120 kV).

*Active microrheology (including sample preparation)*

The magnetic wire microrheology technique has been described in previous accounts.[43,45,57] In brief, wires were synthesized by electrostatic co-assembly of 6.7 nm iron oxide NPs and with poly(diallyldimethylammonium chloride) (PDADMAC, Aldrich, $M_w > 100000$ g mol$^{-1}$)[67] (Supplementary Information S10). The magnetic wires used in this study have lengths between 5 and 100 µm and diameters between 0.6 and 2 µm. The wire diameters were determined independently by optical microscopy and scanning electron microscopy, leading to the expression, $D(L) = 0.619L^{0.202}$ (Supplementary Information S11). As already mentioned, the NP-lipid dispersions were prepared at $c_{PL} = 44$ g L$^{-1}$ and $c_{NP} = 10^{-3}$ and $0.50$ g L$^{-1}$. A volume of 0.5 µL containing $10^5$ wires in PBS was then added to the previous dispersion and gently stirred. 25 µL of this dispersion were then deposited on a glass plate and sealed into a Gene Frame® (Abgene/Advanced Biotech, dimensions $10 \times 10 \times 0.25$ mm³). The glass plate was introduced into a homemade device generating a rotational magnetic field, thanks to two pairs of coils (23 Ω) working with a 90°-phase shift. An electronic set-up allowed measurements in the frequency range $\omega = 10^{-3} - 10^2$ rad s⁻¹ and at magnetic fields $\mu_0 H = 0 - 15$ mTesla. The microrheology protocol used is based on the Magnetic Rotational Spectroscopy technique (Supplementary Information S12).[44-46,57] For each condition of magnetic field and angular frequency, a movie was recorded for a period of time of $10/\omega$ or longer and then treated using the ImageJ software (https://imagej.nih.gov/ij/). For calibration, MRS was performed on a series of water-glycerol mixtures of increasing viscosities, 4.95, 34.9, 48.9 and 80.0 mPa s, corresponding to glycerol concentrations of 49.8%, 81.0%, 84.5% and 89% (T = 25 °C), leading to a susceptibility anisotropy coefficient $\Delta\chi = 0.054 \pm 0.006$ (Supplementary Information S13).[36]

## Acknowledgments

We thank Armelle Baeza-Squiban, Victor Baldim, Yong Chen, Marcel Filoche, Mélody Merle, Mostafa Mokhtari and Chloé Puisney for fruitful discussions. Imane Boucema is acknowledged for letting us use the Anton Paar rheometer for the cone-and-plate rheology. We also thank Stéphane Mornet from the Institut de Chimie de la Matière Condensée de Bordeaux (Université Bordeaux 1) for the synthesis of the aminated silica nanoparticles. ANR (Agence Nationale de la Recherche) and CGI (Commissariat à l'Investissement d'Avenir) are gratefully acknowledged for their financial support of this work through Labex SEAM (Science and Engineering for Advanced Materials and devices) ANR 11 LABX 086, ANR 11 IDEX 05 02. We acknowledge the ImagoSeine facility (Jacques Monod Institute, Paris, France), and the France BioImaging infrastructure supported by the French National Research Agency (ANR-10-INSB-04, « Investments for the future »). This research was supported in part by the Agence Nationale de la Recherche under the contract ANR-13-BS08-0015 (PANORAMA), ANR-12-CHEX-0011 (PULMONANO), ANR-15-CE18-0024-01 (ICONS), ANR-17-CE09-0017 (AlveolusMimics) and by Solvay.

## Supplementary Information

The Supporting Information is available free of charge on the ACS Publications website at DOI: 10.1021/acsnano/xxxxxx

S1: Curosurf® composition and comparison with the native surfactant – S2: Fluid-to-Gel transition from the Curosurf® phospholipid bilayer studied using by differential scanning calorimetry – S3: Dynamic light scattering on Curosurf®/negative silica NPs showing the absence of interaction between the two species – S4: Additional cryo-TEM images of Curosurf® vesicles – S5: Additional oscillation amplitude data $\theta_B$ obtained for Curosurf® loaded with alumina particles – S6: Transmitted light intensity for Curosurf® without and with NPs measured by phase contrast optical microscopy – S7: Mircrorheology data for Curosurf® treated with negatively charged silica – S8: List of production volumes and industrial applications for silica and alumina nanoparticles for 2012 – S9: Curosurf® viscosity as a function of the vesicle volume fraction – S10: Correspondence between the NP concentration in Curosurf® and the amount of nanoparticles in the alveolar region – S11: Wire synthesis scheme – S12: Magnetic wire characterization using optical

microscopy – S13: Magnetic field rotating device and spatial distribution of the magnetic field – S14: Calibrating the magnetic wires using water-glycerol solutions of different viscosity, leading to the determination of the magnetic susceptibility anisotropy.

## TOC Image

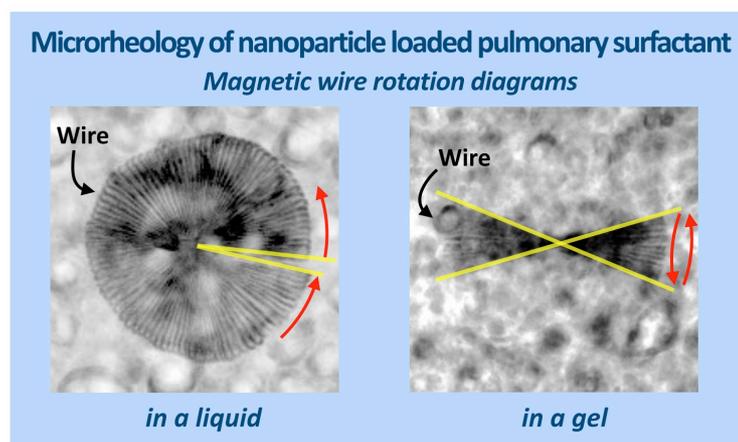

References

1. Falahati, M.; Attar, F.; Sharifi, M.; Haertlé, T.; Berret, J.-F.; Khan, R. H.; Saboury, A. A., A Health Concern Regarding the Protein Corona, Aggregation and Disaggregation. *Biochim. Biophys. Acta Gen. Subjects* **2019**, *1863*, 971 - 991.

2. Moore, T. L.; Rodriguez-Lorenzo, L.; Hirsch, V.; Balog, S.; Urban, D.; Jud, C.; Rothen-Rutishauser, B.; Lattuada, M.; Petri-Fink, A., Nanoparticle Colloidal Stability in Cell Culture Media and Impact on Cellular Interactions. *Chem. Soc. Rev.* **2015**, *44*, 6287-6305.

3. Lundqvist, M.; Stigler, J.; Elia, G.; Lynch, I.; Cedervall, T.; Dawson, K. A., Nanoparticle Size and Surface Properties Determine the Protein Corona with Possible Implications for Biological Impacts. *Proc. Natl. Acad. Sci. U. S. A.* **2008**, *105*, 14265-14270.

4. Walczyk, D.; Bombelli, F. B.; Monopoli, M. P.; Lynch, I.; Dawson, K. A., What the Cell "Sees" in Bionanoscience. *J. Am. Chem. Soc.* **2010**, *132*, 5761-5768.

5. Casals, C.; Canadas, O., Role of Lipid Ordered/Disordered Phase Coexistence in Pulmonary Surfactant Function. *Biochim. Biophys. Acta Biomembranes* **2012**, *1818*, 2550-2562.

6. Lopez-Rodriguez, E.; Perez-Gil, J., Structure-Function Relationships in Pulmonary Surfactant Membranes: From Biophysics to Therapy. *Biochim. Biophys. Acta Biomembranes* **2014**, *1838*, 1568-1585.

7. Richter, R. P.; Berat, R.; Brisson, A. R., Formation of Solid-Supported Lipid Bilayers: An Integrated View. *Langmuir* **2006**, *22*, 3497-3505.

8. Mousseau, F.; Puisney, C.; Mornet, S.; Le Borgne, R.; Vacher, A.; Airiau, M.; Baeza-Squiban, A.; Berret, J. F., Supported Pulmonary Surfactant Bilayers on Silica Nanoparticles: Formulation, Stability and Impact on Lung Epithelial Cells. *Nanoscale* **2017**, *9*, 14967-14978.

9. Liu, J. W., Interfacing Zwitterionic Liposomes with Inorganic Nanomaterials: Surface Forces, Membrane Integrity, and Applications. *Langmuir* **2016**, *32*, 4393-4404.

10. Mornet, S.; Lambert, O.; Duguet, E.; Brisson, A., The Formation of Supported Lipid Bilayers on Silica Nanoparticles Revealed by Cryoelectron Microscopy. *Nano Lett.* **2005**, *5*, 281-285.

11. Michel, R.; Gradzielski, M., Experimental Aspects of Colloidal Interactions in Mixed Systems of Liposome and Inorganic Nanoparticle and Their Applications. *Int. J. Mol. Sci.* **2012**, *13*, 11610-11642.

12.     Dasgupta, S.; Auth, T.; Gov, N. S.; Satchwell, T. J.; Hanssen, E.; Zuccala, E. S.; Riglar, D. T.; Toye, A. M.; Betz, T.; Baum, J.; Gompper, G., Membrane-Wrapping Contributions to Malaria Parasite Invasion of the Human Erythrocyte. *Biophys. J.* **2014**, *107*, 43-54.

13.     Deserno, M.; Gelbart, W. M., Adhesion and Wrapping in Colloid-Vesicle Complexes. *J. Phys. Chem. B* **2002**, *106*, 5543-5552.

14.     Froehlich, E., The Role of Surface Charge in Cellular Uptake and Cytotoxicity of Medical Nanoparticles. *Int. J. Nanomed.* **2012**, *7*, 5577-5591.

15.     Hidalgo, A.; Cruz, A.; Perez-Gil, J., Pulmonary Surfactant and Nanocarriers: Toxicity *versus* Combined Nanomedical Applications. *Biochim. Biophys. Acta Biomembranes* **2017**, *1859*, 1740-1748.

16.     Hobi, N.; Siber, G.; Bouzas, V.; Ravasio, A.; Perez-Gil, J.; Haller, T., Physiological Variables Affecting Surface Film Formation by Native Lamellar Body-Like Pulmonary Surfactant Particles. *Biochim. Biophys. Acta Biomembranes* **2014**, *1838*, 1842-1850.

17.     Lewis, J. F.; Jobe, A. H., Surfactant and the Adult Respiratory-Distress Syndrome. *Am. Rev. Respir. Dis.* **1993**, *147*, 218-233.

18.     Numata, M.; Kandasamy, P.; Voelker, D. R., Anionic Pulmonary Surfactant Lipid Regulation of Innate Immunity. *Expert Rev. Respir. Med.* **2012**, *6*, 243-246.

19.     Notter, R. H., *Lung Surfactant: Basic Science and Clinical Applications*. CRC Press: Boca Raton, FL, 2000; Vol. 149.

20.     Gil, J.; Weibel, E. R., Morphological Study of Pressure-Volume Hysteresis in Rat Lungs Fixed by Vascular Perfusion. *Resp. Physiol.* **1972**, *15*, 190-213.

21.     Schleh, C.; Muhlfeld, C.; Pulskamp, K.; Schmiedl, A.; Nassimi, M.; Lauenstein, H. D.; Braun, A.; Krug, N.; Erpenbeck, V. J.; Hohlfeld, J. M., The Effect of Titanium Dioxide Nanoparticles on Pulmonary Surfactant Function and Ultrastructure. *Respir. Res.* **2009**, *10*, 90.

22.     Beck-Broichsitter, M.; Ruppert, C.; Schmehl, T.; Guenther, A.; Betz, T.; Bakowsky, U.; Seeger, W.; Kissel, T.; Gessler, T., Biophysical Investigation of Pulmonary Surfactant Surface Properties upon Contact with Polymeric Nanoparticles *In Vitro. Nanomedicine* **2011**, *7*, 341-350.

23.     Pera, H.; Nolte, T. M.; Leermakers, F. A. M.; Kleijn, J. M., Coverage and Disruption of Phospholipid Membranes by Oxide Nanoparticles. *Langmuir* **2014**, *30*, 14581-14590.

24.     Bakshi, M. S.; Zhao, L.; Smith, R.; Possmayer, F.; Petersen, N. O., Metal Nanoparticle Pollutants Interfere with Pulmonary Surfactant Function *In Vitro. Biophys. J.* **2008**, *94*, 855-868.

25.     Harishchandra, R. K.; Saleem, M.; Galla, H.-J., Nanoparticle Interaction with Model Lung Surfactant Monolayers. *J. R. Soc. Interface* **2010**, *7*, S15-S26.

26.     Sachan, A. K.; Harishchandra, R. K.; Bantz, C.; Maskos, M.; Reichelt, R.; Galla, H. J., High-Resolution Investigation of Nanoparticle Interaction with a Model Pulmonary Surfactant Monolayer. *ACS Nano* **2012**, *6*, 1677-1687.

27.     Fan, Q.; Wang, Y. E.; Zhao, X.; Loo, J. S. C.; Zuo, Y. Y., Adverse Biophysical Effects of Hydroxyapatite Nanoparticles on Natural Pulmonary Surfactant. *ACS Nano* **2011**, *5*, 6410-6416.

28.     Kodama, A. T.; Kuo, C.-C.; Boatwright, T.; Dennin, M., Investigating the Effect of Particle Size on Pulmonary Surfactant Phase Behavior. *Biophys. J.* **2014**, *107*, 1573-1581.

29.     Kondej, D.; Sosnowski, T. R., Effect of Clay Nanoparticles on Model Lung Surfactant: A Potential Marker of Hazard from Nanoaerosol Inhalation. *Environ. Sci. Poll. Res.* **2016**, *23*, 4660-4669.

30.     Xia, T.; Zhu, Y.; Mu, L.; Zhang, Z.-F.; Liu, S., Pulmonary Diseases Induced by Ambient Ultrafine and Engineered Nanoparticles in Twenty-First Century. *Nation. Sci. Rev.* **2016**, *3*, 416-429.

31.     Lelieveld, J.; Evans, J. S.; Fnais, M.; Giannadaki, D.; Pozzer, A., The Contribution of Outdoor Air Pollution Sources to Premature Mortality on a Global Scale. *Nature* **2015**, *525*, 367.

32.     King, D. M.; Wang, Z. D.; Palmer, H. J.; Holm, B. A.; Notter, R. H., Bulk Shear Viscosities of Endogenous and Exogenous Lung Surfactants. *Am. J. Physiol. Lung Cell Mol. Physiol.* **2002**, *282*, L277-L284.

33.     Lu, K. W.; Perez-Gil, J.; Taeusch, H. W., Kinematic Viscosity of Therapeutic Pulmonary Surfactants with Added Polymers. *Biochim. Biophys. Acta Biomembranes* **2009**, *1788*, 632-637.

34.     Krieger, I. M.; Dougherty, T. J., A Mechanism for Non-Newtonian Flow in Suspensions of Rigid Spheres. *Trans. Soc. Rheol.* **1959**, *3*, 137-152.

35.     Larson, R. G., *The Structure and Rheology of Complex Fluids*. Oxford University Press: New York, 1998.

36.     Thai, L. P. A.; Mousseau, F.; Oikonomou, E. K.; Berret, J. F., On the Rheology of Pulmonary Surfactant: Effects of Concentration and Consequences for the Surfactant Replacement Therapy. *Colloids Surf. B: Biointerfaces* **2019**, *178*, 337-345.

37.     van Donkelaar, A.; Martin, R. V.; Brauer, M.; Kahn, R.; Levy, R.; Verduzco, C.; Villeneuve, P. J., Global Estimates of Ambient Fine Particulate Matter Concentrations from Satellite-Based Aerosol Optical Depth: Development and Application. *Environ. Health Perspect.* **2010**, *118*, 847-855.

38.     Lanone, S.; Rogerieux, F.; Geys, J.; Dupont, A.; Maillot-Marechal, E.; Boczkowski, J.; Lacroix, G.; Hoet, P., Comparative Toxicity of 24 Manufactured Nanoparticles in Human Alveolar Epithelial and Macrophage Cell Lines. *Part. Fibre Toxicol.* **2009**, *6*, 14.

39.     Napierska, D.; Thomassen, L. C. J.; Lison, D.; Martens, J. A.; Hoet, P. H., The Nanosilica Hazard: Another Variable Entity. *Part. Fibre Toxicol.* **2010**, *7*, 39.

40.     Oberdorster, G.; Oberdorster, E.; Oberdorster, J., Nanotoxicology: An emerging discipline evolving from studies of ultrafine particles. *Environ. Health Perspect.* **2005**, *113*, 823-839.

41.     Puisney, C.; Oikonomou, E. K.; Nowak, S.; Chevillot, A.; Casale, S.; Baeza-Squiban, A.; Berret, J.-F., Brake Wear (Nano)Particle Characterization and Toxicity on Airway Epithelial Cells *In Vitro*. *Environ. Sci. Nano* **2018**, *5*, 1036-1044.

42.     Geiser, M.; Kreyling, W. G., Deposition and biokinetics of inhaled nanoparticles. *Part. Fibre Toxicol.* **2010**, *7*, 2.

43.     Berret, J.-F., Local Viscoelasticity of Living Cells Measured by Rotational Magnetic Spectroscopy. *Nat. Commun.* **2016**, *7*, 10134.

44.     Gu, Y.; Kornev, K. G., Ferromagnetic Nanorods in Applications to Control of the In-Plane Anisotropy of Composite Films and for *In Situ* Characterization of the Film Rheology. *Adv. Funct. Mater.* **2016**, *26*, 3796-3808.

45.     Loosli, F.; Najm, M.; Chan, R.; Oikonomou, E.; Grados, A.; Receveur, M.; Berret, J.-F., Wire-Active Microrheology to Differentiate Viscoelastic Liquids from Soft Solids. *ChemPhysChem* **2016**, *17*, 4134-4143.

46.     Tokarev, A.; Yatvin, J.; Trotsenko, O.; Locklin, J.; Minko, S., Nanostructured Soft Matter with Magnetic Nanoparticles. *Adv. Funct. Mater.* **2016**, *26*, 3761-3782.

47.     Mousseau, F.; Berret, J. F., The Role of Surface Charge in the Interaction of Nanoparticles with Model Pulmonary Surfactants. *Soft Matter* **2018**, *14*, 5764-5774.

48.     Mousseau, F.; Le Borgne, R.; Seyrek, E.; Berret, J.-F., Biophysicochemical Interaction of a Clinical Pulmonary Surfactant with Nanoalumina. *Langmuir* **2015**, *31*, 7346-7354.

49.     Mousseau, F.; Vitorazi, L.; Herrmann, L.; Mornet, S.; Berret, J. F., Polyelectrolyte Assisted Charge Titration Spectrometry: Applications to Latex and Oxide Nanoparticles. *J. Colloid Interface Sci.* **2016**, *475*, 36-45.

50.     Braun, A.; Stenger, P. C.; Warriner, H. E.; Zasadzinski, J. A.; Lu, K. W.; Taeusch, H. W., A Freeze-Fracture Transmission Electron Microscopy and Small Angle X-Ray Diffraction Study of the Effects of Albumin, Serum, and Polymers on Clinical Lung Surfactant Microstructure. *Biophys. J.* **2007**, *93*, 123-139.

51.     Kumar, A.; Bicer, E. M.; Morgan, A. B.; Pfeffer, P. E.; Monopoli, M.; Dawson, K. A.; Eriksson, J.; Edwards, K.; Lynham, S.; Arno, M.; Behndig, A. F.; Blomberg, A.; Somers, G.; Hassall, D.; Dailey, L. A.; Forbes, B.; Mudway, I. S., Enrichment of Immunoregulatory Proteins in the Biomolecular Corona of Nanoparticles within Human Respiratory Tract Lining Fluid. *Nanomedicine* **2016**, *12*, 1033-1043.

52.     De Backer, L.; Braeckmans, K.; Stuart, M. C. A.; Demeester, J.; De Smedt, S. C.; Raemdonck, K., Bio-inspired pulmonary surfactant-modified nanogels: A promising siRNA delivery system. *J. Control. Release* **2015**, *206*, 177-186.

53.     Waisman, D.; Danino, D.; Weintraub, Z.; Schmidt, J.; Talmon, Y., Nanostructure of the Aqueous Form of Lung Surfactant of Different Species Visualized by Cryo-Transmission Electron Microscopy. *Clin. Physiol. Funct. Imaging* **2007**, *27*, 375-380.

54.     Belloni, L., Ionic Condensation and Charge Renormalization in Colloid Suspensions. *Colloids Surf. A* **1998**, *140*, 227 - 243.

55. Job, P., Studies on the Formation of Complex Minerals in Solution and on Their Stability. *Ann. Chim. France* **1928,** *9*, 113-203.

56. Renny, J. S.; Tomasevich, L. L.; Tallmadge, E. H.; Collum, D. B., Method of Continuous Variations: Applications of Job Plots to the Study of Molecular Associations in Organometallic Chemistry. *Angew. Chem.-Int. Edit.* **2013,** *52*, 11998-12013.

57. Chevry, L.; Sampathkumar, N. K.; Cebers, A.; Berret, J. F., Magnetic Wire-Based Sensors for the Microrheology of Complex Fluids. *Physical Review E* **2013,** *88*, 062306.

58. Helgesen, G.; Pieranski, P.; Skjeltorp, A. T., Nonlinear Phenomena in Systems of Magnetic Holes. *Phys. Rev. Lett.* **1990,** *64*, 1425-1428.

59. Minami, S.; Watanabe, T.; Suzuki, D.; Urayama, K., Rheological Properties of Suspensions of Thermo-Responsive Poly(N-Isopropylacrylamide) Microgels Undergoing Volume Phase Transition. *Polymer J.* **2016,** *48*, 1079-1086.

60. Varga, Z.; Swan, J. W., Linear Viscoelasticity of Attractive Colloidal Dispersions. *J. Rheol.* **2015,** *59*, 1271-1298.

61. Weis, C.; Oelschlaeger, C.; Dijkstra, D.; Ranft, M.; Willenbacher, N., Microstructure, Local Dynamics, and Flow Behavior of Colloidal Suspensions with Weak Attractive Interactions. *Sci. Rep.* **2016,** *6*, 33498.

62. Weis, C.; Natalia, I.; Willenbacher, N., Effect of Weak Attractive Interactions on Flow Behavior of Highly Concentrated Crystalline Suspensions. *J. Rheol.* **2014,** *58*, 1583-1597.

63. Curstedt, T.; Halliday, H. L.; Speer, C. P., A Unique Story in Neonatal Research: The Development of a Porcine Surfactant. *Neonatology* **2015,** *107*, 321-329.

64. Panaiotov, I.; Ivanova, T.; Proust, J.; Boury, F.; Denizot, B.; Keough, K.; Taneva, S., Effect of Hydrophobic Protein Sp-C on Structure and Dilatational Properties of The Model Monolayers of Pulmonary Surfactant. *Colloids Surf. B: Biointerfaces* **1996,** *6*, 243-260.

65. Camacho, L.; Cruz, A.; Castro, R.; Casals, C.; Pérez-Gil, J., Effect of Ph on the Interfacial Adsorption Activity of Pulmonary Surfactant. *Colloids Surf. B: Biointerfaces* **1996,** *5*, 271-277.

66. Dubochet, J., On the Development of Electron Cryo-Microscopy (Nobel Lecture). *Angew. Chem.-Int. Edit.* **2018,** *57*, 10842-10846.

67. Yan, M.; Fresnais, J.; Sekar, S.; Chapel, J. P.; Berret, J.-F., Magnetic Nanowires Generated *via* the Waterborne Desalting Transition Pathway. *ACS Appl. Mater. Interfaces* **2011,** *3*, 1049-1054.